\documentclass[twocolumn]{openjournal}
\usepackage{natbib}
\usepackage{graphicx,amsmath,amssymb,amstext}
\usepackage{amsbsy,amsfonts,amsthm,color}
\usepackage{chemformula}
\usepackage[colorlinks,linkcolor=blue,citecolor=blue,urlcolor=blue ]{hyperref}
\usepackage[utf8]{inputenc}
\usepackage{float}
\usepackage[caption=false]{subfig}
\newcommand{\pcc}{\,{\rm cm}^{-3}}
\newcommand{\gcc}{\,{\rm g \, cm}^{-3}}
\newcommand{\kel}{\, {\rm K}}
\newcommand{\msun}{\, {\rm M}_\odot}

\newcommand{\pc}{\, {\rm pc}}
\newcommand{\myr}{\, {\rm Myr}}
\newcommand{\co}{\ch{CO}}
\newcommand{\cs}{\ch{CS}}
\newcommand{\hco}{\ch{HCO+}}
\newcommand{\nth}{\ch{N2H+}}
\newcommand{\kms}{\, {\rm km \, s^{-1}}}

\defcitealias{Prole2023}{LP23}
\defcitealias{Schauer2021}{AS21}

\begin{document}

%\DeclareSIUnit \parsec {pc}

\title{Cloud Collision Signatures in the Central Molecular Zone\vspace{-3em}}

\author{Rees A. Barnes$^{* \ 1,2}$}
\author{Felix D. Priestley$^{1}$}
\email{$^*$email: rb5132@princeton.edu}

\affiliation {School of Physics and Astronomy, Cardiff University, Queen’s Buildings, The Parade, Cardiff CF24 3AA, UK}
\affiliation {Department of Physics, Princeton University, Princeton, NJ, USA}

\date{July 2024}

\begin{abstract}

\noindent Molecular cloud collisions are a prominent theory for the formation of stars.  Observational studies into cloud collisions identify the collision via a bridging feature: a continuous strip of line emission that connects two intensity peaks that are related in position space and separated in velocity space.  Currently, most observations of collisions and these bridging features take place in the Milky Way disc.  They are also theorized to take place in the Central Molecular Zone (CMZ), where temperatures and densities are both significantly higher than in the disc.  For studies in the Milky Way Disc, the most commonly-used tracer tends to be \( \co \).  However, for studies in the CMZ, where the density and temperature are significantly higher, low-J \( \co \) lines lose their ability to adequately highlight the bridging feature of cloud collisions.  As a result, studies have begun using other tracers, whose physical and chemical behavior has not been studied under CMZ conditions. We perform combined hydrodynamical, chemical and radiative transfer simulations of cloud collisions under both disc- and CMZ-like conditions, and investigate collision signatures in a number of commonly-observed molecular lines.  Under the Milky Way disc conditions \( \co \) has the standard bridging feature; however, the other tracers, \( \cs \), \( \hco \), \( \nth \) only emit in the intermediate-velocity bridge region, making the feature itself challenging to detect.  In the CMZ, the higher density and temperature make the bridging feature far more indistinct for \( \co \), but the other tracers have morphologically similar bridging features to the \( \co \) disc model, validating their use as tracers of cloud collisions under these conditions.

%After examining \( \co \), \( \cs \), \( \hco \), and \( \nth \) in the Milky Way Disc, CMZ, and CMZ with a high ionization rate environments, we have concluded that in the CMZ \( \cs \) proved to be the best and most valuable tracer.  In the CMZ with a high ionization rate, \( \hco \) and \( \nth \) proved to be the most suitable tracer molecules.

\end{abstract}


\maketitle

\section{Introduction} 
\label{sec:intro}
%Finding the Population III (Pop III) initial mass function (IMF) remains a sought after goal for use in high volume cosmological simulations, which can not afford the high resolutions attainable in small-scale simulations due to the vast scope of the simulations. These cosmological simulations explore mechanisms such as the reionisation of the Universe (e.g. \citealt{Rosdahl2018,Garel2021}), metal enrichment and mixing (e.g. \citealt{Jaacks2018,Magg2022}), black hole formation (e.g. \citealt{Smith2018,Dave2019}), direct collapse black hole formation (e.g. \citealt{Schauer2017, Chon2021a}, reproducing observations (e.g. \citealt{Bolton2017}) etc. In the absence of sufficient resolution to resolve individual star formation, a prescription for the IMF within primordial halos is required once the mesh reaches its highest refinement level.

%Despite this, small-scale simulations investigating singular star forming halos are able to either follow the IMF for 10$^3$-10$^4$ yr with insufficient resolution to resolve the protostar  (e.g. \citealt{Stacy2013,Susa2014,Stacy2016,Wollenberg2020,Sharda2020,Jaura2022}) or for considerably less time  (10$^2$-10$^3$ yr) using protostellar resolutions (e.g. \citealt{Greif2012,Prole2022,Prole2022a,Hirano2022,Prole2023}). Studies have shown that using lower resolutions underestimates fragmentation within the gas and hence overestimates stellar masses through the absence of fragmentation induced starvation \citep{Peters2010,Machida2013,Prole2022}.

\noindent Current theory suggests that collisions between molecular clouds are potential triggering mechanisms for star formation \citep{tasker2009} and are expected to be fairly common occurrences in galactic discs \citep{dobbs2015}.  These cloud collisions have also been observed outside of the Milky Way's disk, in the Small and Large Magellanic Clouds \citep{fukui2019,tokuda2019,neelamkodan2021} and are theorized to take place closer to the galactic center, as gas flows into the Central Molecular Zone (CMZ) \citep{sormani2019}.

When searching for high velocity molecular cloud collisions in observational data, the most common method of identification is via \( \co \) bridging features \citep{haworth2015}.  This signature consists of two intensity peaks in position-velocity space connected by intermediate-velocity emission, corresponding to the material decelerated by the collision.  Cloud collision signatures, usually bridging features, were identified in more than 50 high-mass star formation regions of the Milky Way disc \citep{fukui2021}.

In the Milky Way disc, these bridging features are identified with CO as a tracer. Recent studies of the CMZ have, in addition to \( \co \) \citep{enokiya2021}, claimed to identify bridging features in lines of other species such as \( \cs \) \citep{busch2022} and \ch{SiO} \citep{armijos2020}.  \( \cs \), being a dense gas tracer, should not show any bridging features under the normal Milky Way conditions as the emission is concentrated in the density-enhanced intermediate velocity gas \citep{priestley2021}.  In the Milky Way disc, SiO only traces shocked gas and as such should only present in the bridge region itself, rather than tracing the entire bridge feature \citep{consentino2020}. Within the CMZ, however, the environment behaves entirely different from the normal Milky Way conditions of previous bridging feature studies.  As such, the common methods of identifying bridging features may not be appropriate for this environment.  Observational studies suggest that molecular clouds in the CMZ are denser, hotter, and more turbulent than the Milky Way disc \citep{henshaw2023}.  Furthermore, these regions of the CMZ have a higher ionization rate, which may have a significant impact on their chemistry \citep{indriolo2015}.

This paper seeks to examine observational signatures of cloud collisions under CMZ conditions via a combination of hydrodynamical, chemical, and radiative transfer models.  The distinct chemical properties of the CMZ \citep{henshaw2019,santamaria2021} require that the molecular abundances are evolved self-consistently, as simplifications such as spatially and temporally invariant abundances are likely to produce erroneous results even under better-understood Milky Way disc conditions \citep{priestley2023, clement2023}.  Our synthetic observations will allow us to investigate the validity of extrapolating the well-known \( \co \) bridging feature signature into a regime which has not previously been studied.

\ \
\section{Method}
\label{sec:method}

%\begin{figure*}
	% To include a figure from a file named example.*
	% Allowable file formats are eps or ps if compiling using latex
	% or pdf, png, jpg if compiling using pdflatex
	% \hbox{\hspace{0cm} \includegraphics[scale=0.7]{barotropic_EoS.pdf}}
   Our simulation setup is similar to previous studies on this topic \citep{haworth2015,priestley2021}, and consists of a head-on collision between two initially-spherical, differently-sized clouds. The larger cloud has an initial radius of $5 \pc$, the smaller one a radius of $2.5 \pc$, and the cloud centres are separated by $10 \pc$ at the beginning of the simulation. Both clouds have an initially uniform density, with the smaller cloud being twice as dense as the larger one. Each cloud is initialised with a turbulent velocity field following \citet{lomax2015}, and given a bulk motion of $10 \kms$ towards each other. We perform simulations using the smoothed-particle hydrodynamics (SPH) code {\sc phantom} \citep{price2018} with $375,000$ particles in the clouds, and a background medium with a density $100$ times lower than the cloud material with the sound speed increased to ensure pressure balance, containing another $\sim 100,000$ particles. Sink particles are introduced with a formation density of $10^{-17} \gcc$ and an accretion radius of $0.01 \pc$.

We perform two SPH simulations, representing conditions typical for the Galactic disc and the CMZ. The disc simulation has a large cloud mass of $10^4 \msun$ and a small cloud mass of $2500 \msun$, giving initial volume densities of $550 \pcc$ and $1100 \pcc$ respectively. The SPH particle mass is $0.03 \msun$. The sound speed within the clouds is set to $0.19 \kms$, appropriate for molecular gas at a temperature of $10 \kel$, and the root-mean-squared velocity of the turbulent velocity field is set to five times this value. For the CMZ simulation, we increase the cloud masses to $10^5 \msun$ and $2.5 \times 10^4 \msun$ and the sound speed to $0.42 \kms$ (i.e. $50 \kel$). We keep all other parameters fixed, so the turbulent velocity field has a higher average value but the same Mach number, the collision speed is unchanged, and the particle mass is now $0.3 \msun$. The simulations are run until they have accreted a few percent of the cloud mass into sink particles, which occurs after $0.32 \myr$ for the disc simulation and $0.18 \myr$ for the CMZ one.
\begin{figure*}
    \centering
    \includegraphics[width=0.3\textwidth]{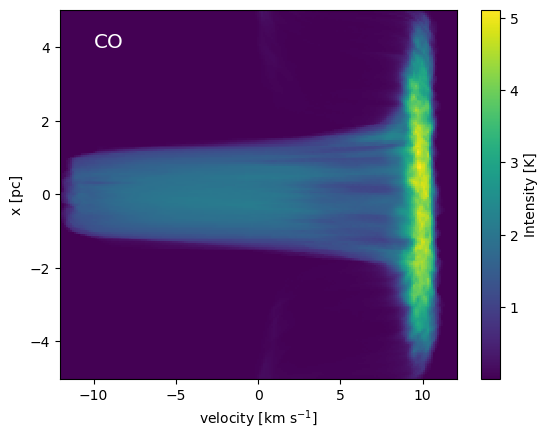}
    \includegraphics[width=0.3\textwidth]{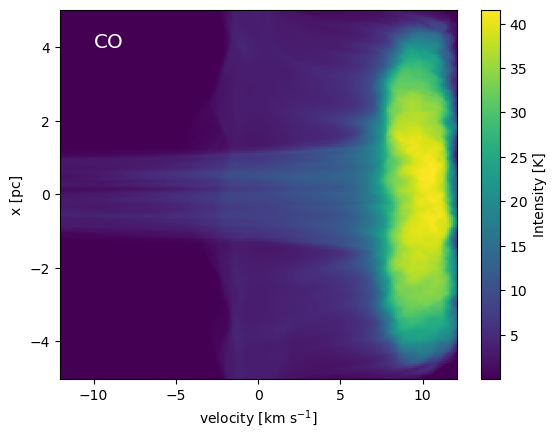}
    \includegraphics[width=0.3\textwidth]{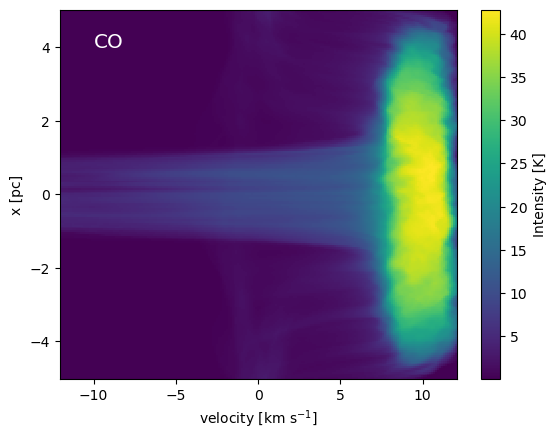}
    \\
    \includegraphics[width=0.3\textwidth]{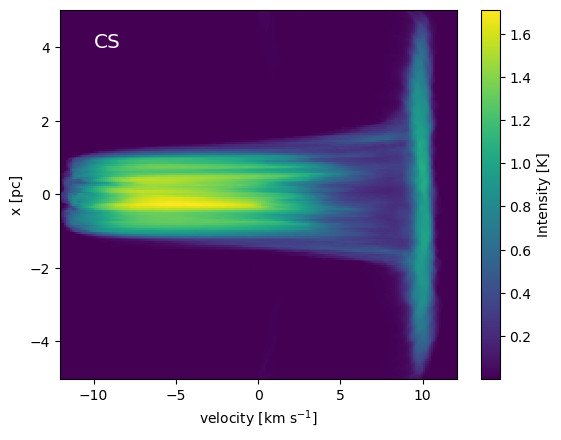}
    \includegraphics[width=0.3\textwidth]{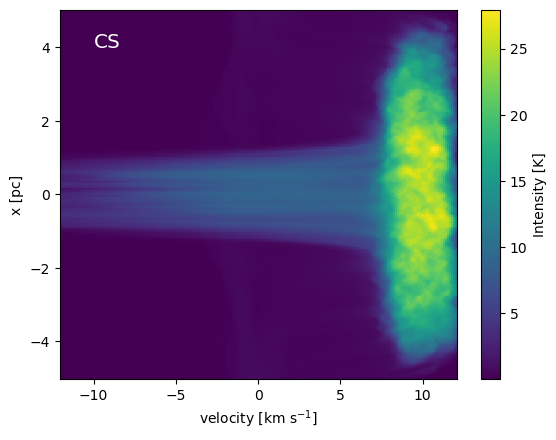}
    \includegraphics[width=0.3\textwidth]{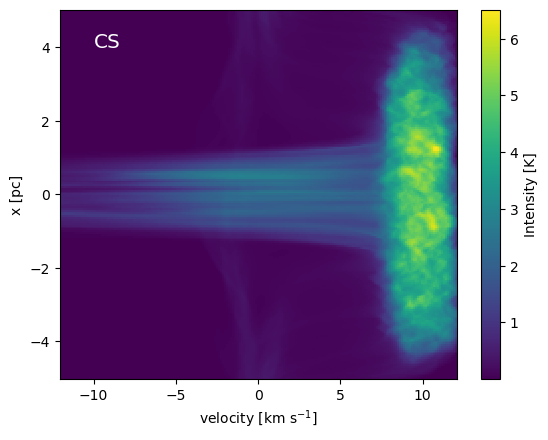}
    \\
    \includegraphics[width=0.3\textwidth]{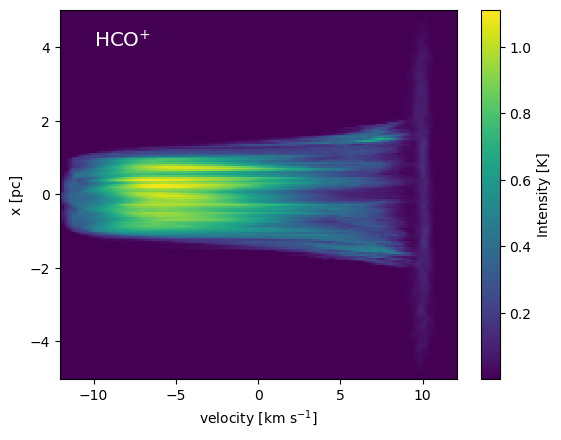}
    \includegraphics[width=0.3\textwidth
    ]{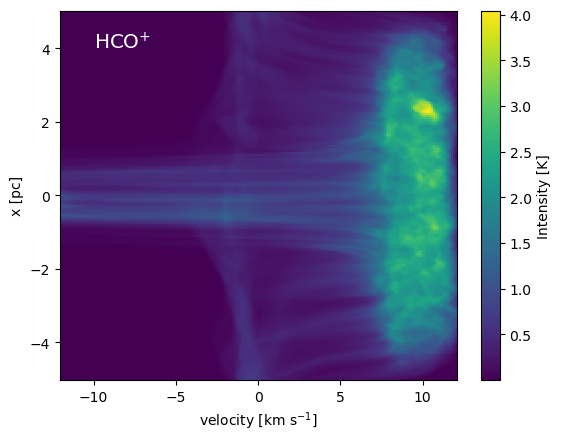}
    \includegraphics[width=0.3\textwidth]{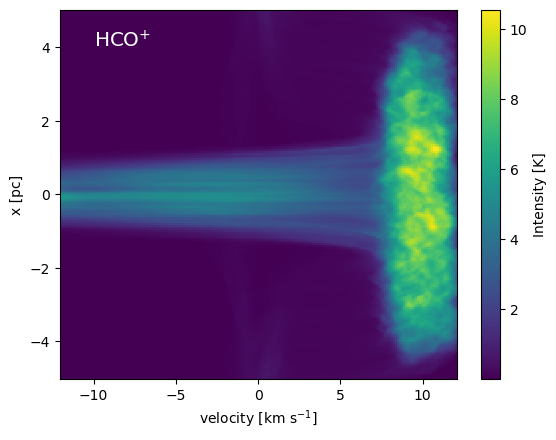}
    \\
    \includegraphics[width=0.3\textwidth]{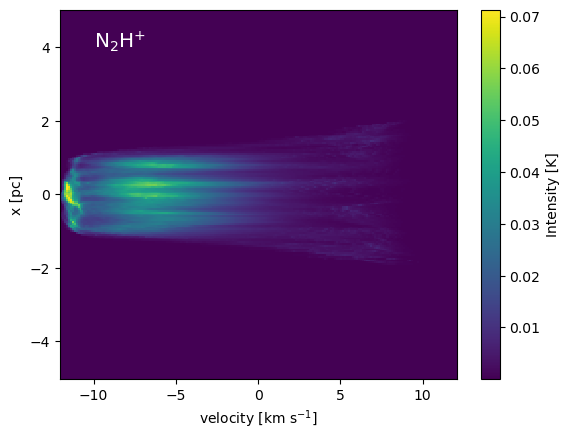}
    \includegraphics[width=0.3\textwidth]{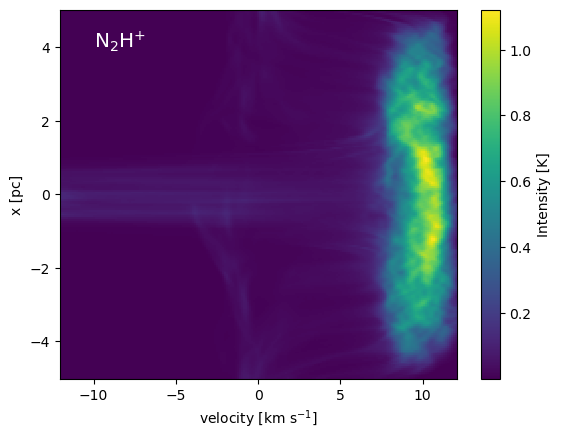}
    \includegraphics[width=0.3\textwidth]{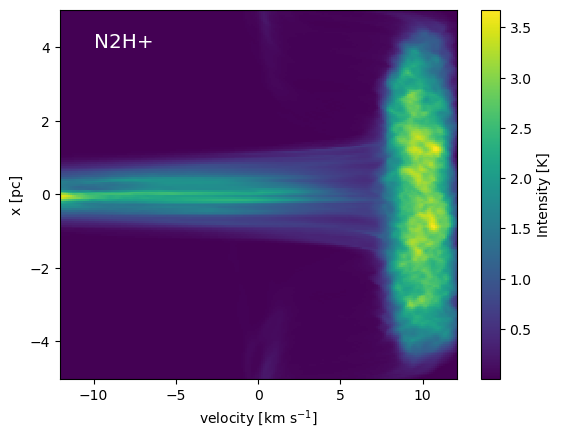}
\caption{Position-Velocity plots (averaged along the y-axis) from the three different simulations. \\ Left Column: Milky Way Disc - Center Column: CMZ - Right Column: CMZ with high $\zeta$}
\label{fig:yAvg}
\end{figure*}

\begin{figure*}
    \centering
    \includegraphics[width=0.3\textwidth]{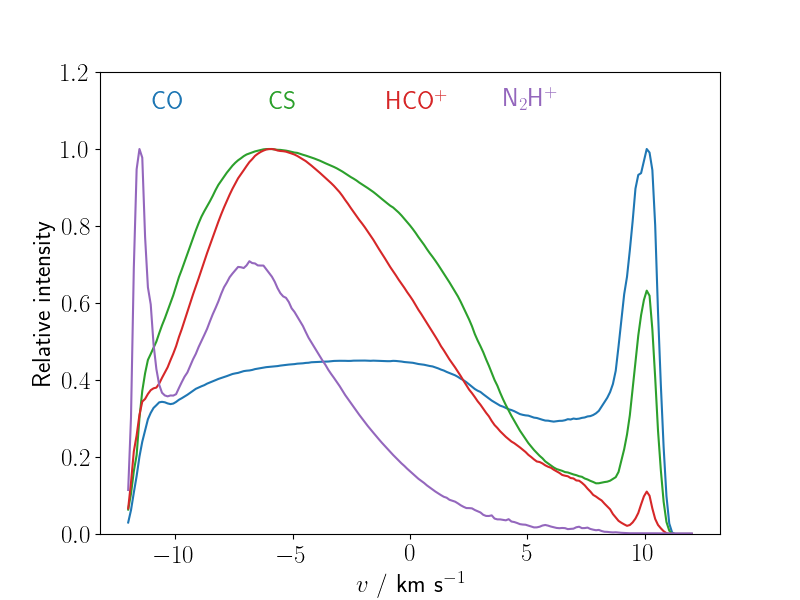}
    \includegraphics[width=0.3\textwidth]{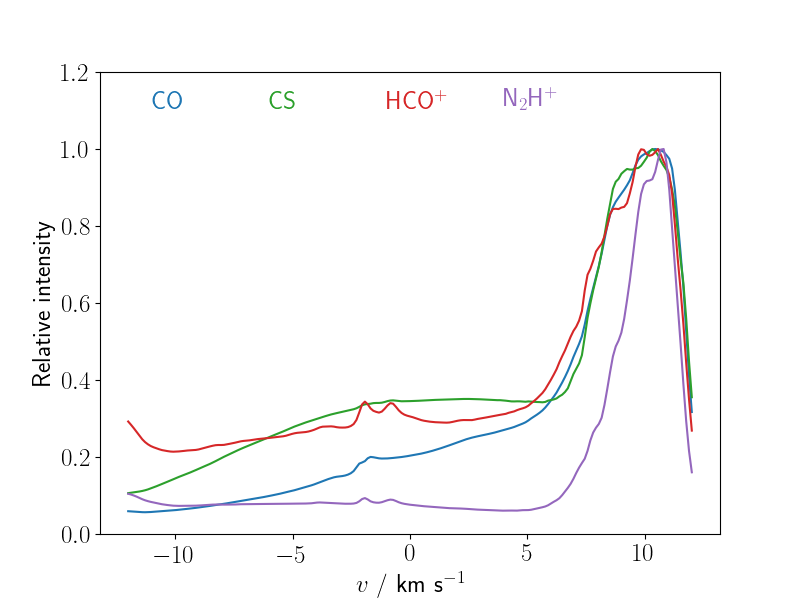}
    \includegraphics[width=0.3\textwidth]{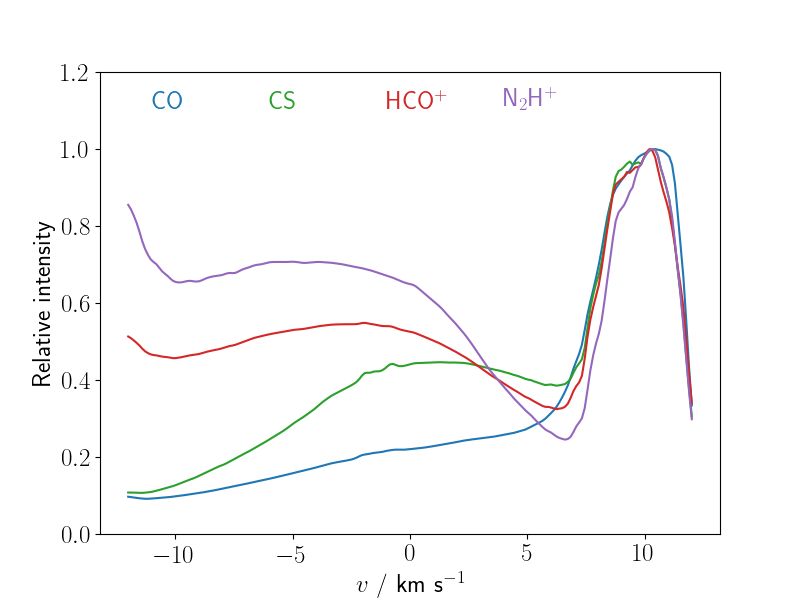}
\caption{Normalised line profiles of \ch{CO}, \ch{CS}, \ch{HCO+}, and \ch{N2H+} averaged over the central $0.5$ pc region. \\ Left: Milky Way Disc - Center: CMZ - Right: CMZ with high $\zeta$}
\label{fig:profile}
\end{figure*}

\begin{figure*}
    \centering
    \includegraphics[width=0.3\textwidth]{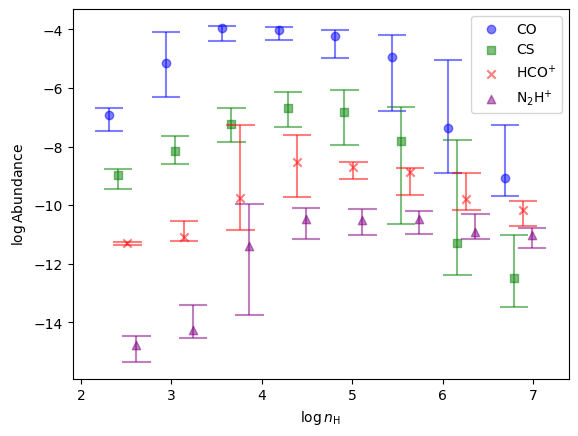}
    \includegraphics[width=0.3\textwidth]{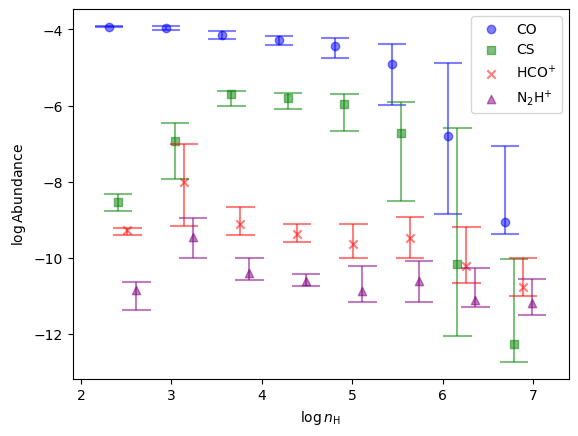}
    \includegraphics[width=0.3\textwidth]{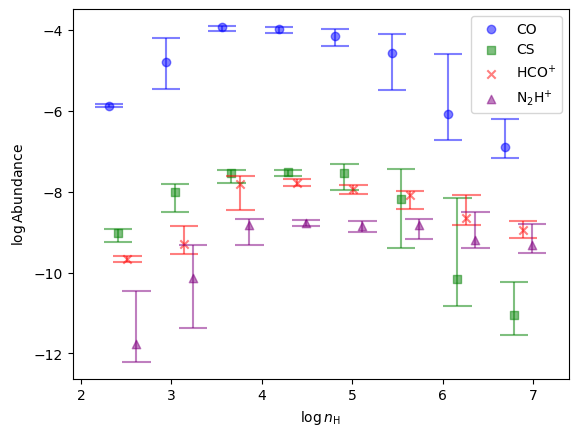}
\caption{Abundance of \ch{CO}, \ch{CS}, \ch{HCO+}, and \ch{N2H+} as a function of Hydrogen Nuclei Density. \\ Left: Milky Way Disc - Center: CMZ - Right: CMZ with high $\zeta$}
\label{fig:envAbun}
\end{figure*}

%\begin{figure*}[h]
   %\centering
   %\includegraphics[width=0.4\textwidth]{PaperGraphs/abundance/co_abun.png}
   %\includegraphics[width=0.4\textwidth]{PaperGraphs/abundance/cs_abun.png}
   %\\
   %\includegraphics[width=0.4\textwidth]{PaperGraphs/abundance/hco+_abun.png}
   %\includegraphics[width=0.4\textwidth]{PaperGraphs/abundance/n2h+_abun.png}
%\caption{Abundance of Molecules in each environment}
%\label{fig:molAbun}
%\end{figure*}

\begin{figure*}
    \centering
    \includegraphics[width=0.3\textwidth]{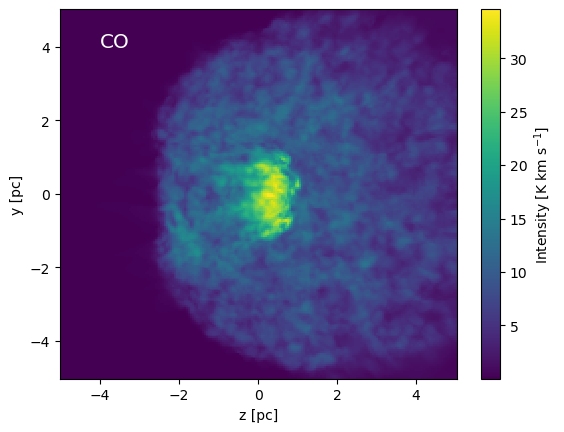}
    \includegraphics[width=0.3\textwidth]{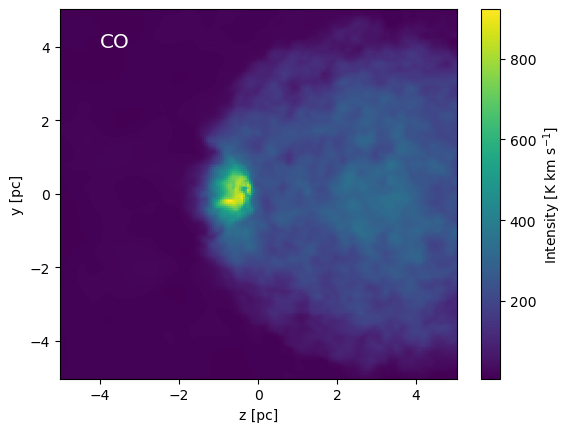}
    \includegraphics[width=0.3\textwidth]{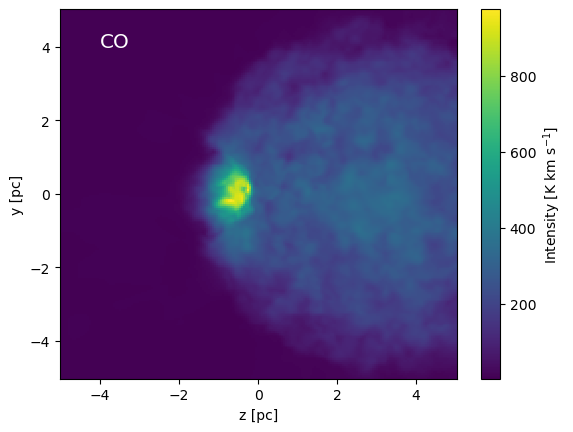}
    \\
    \includegraphics[width=0.3\textwidth]{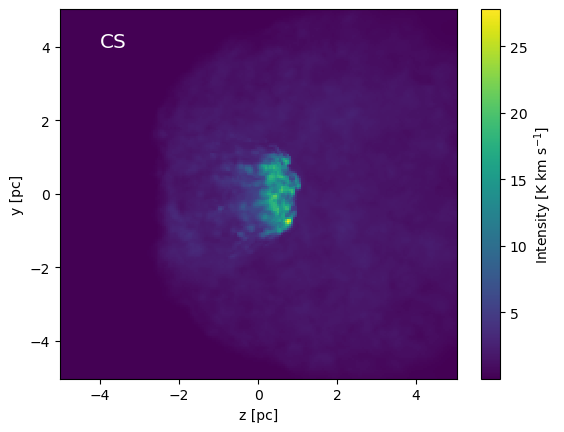}
    \includegraphics[width=0.3\textwidth]{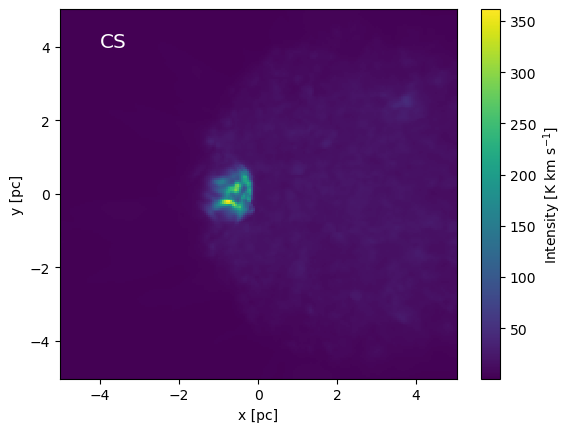}
    \includegraphics[width=0.3\textwidth]{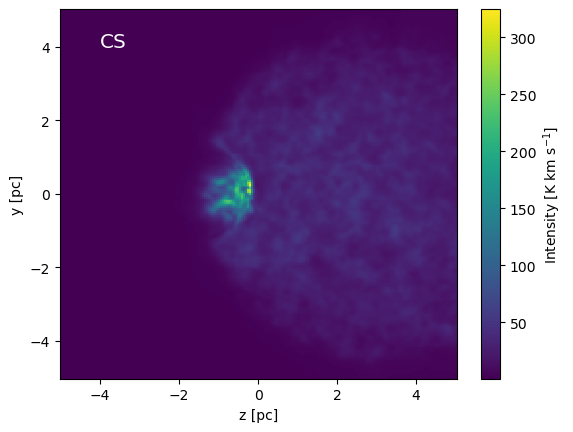}
    \\
    \includegraphics[width=0.3\textwidth]{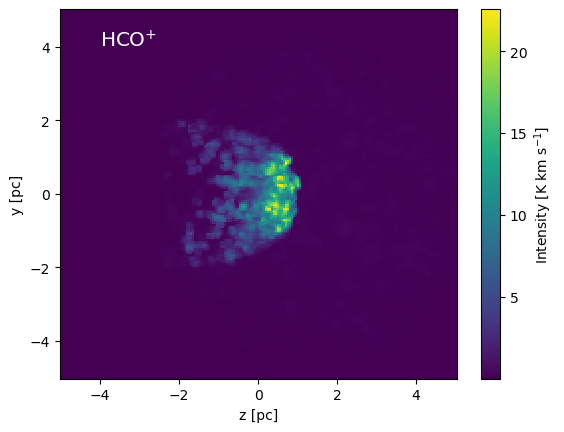}
    \includegraphics[width=0.3\textwidth]{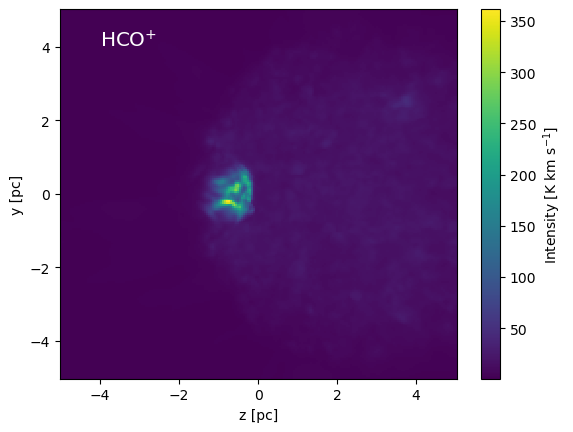}
    \includegraphics[width=0.3\textwidth]{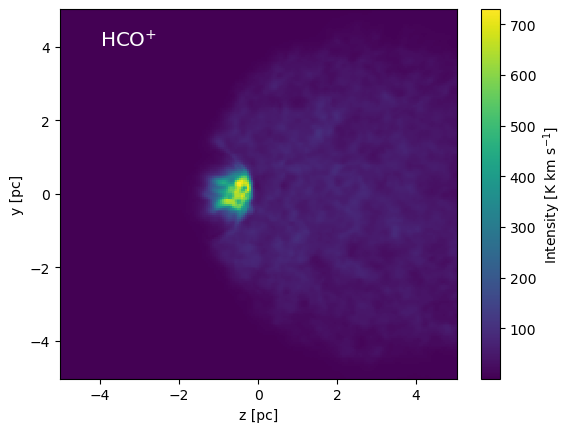}
    \\
    \includegraphics[width=0.3\textwidth]{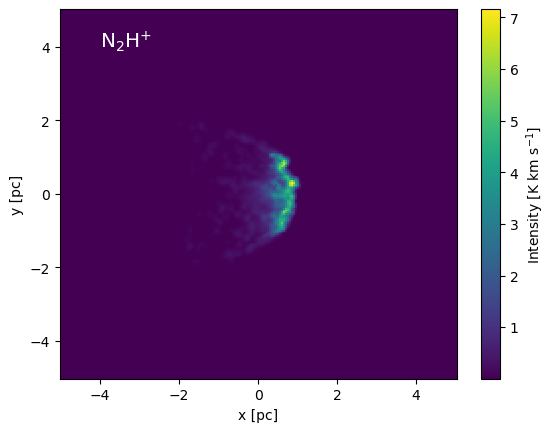}
    \includegraphics[width=0.3\textwidth]{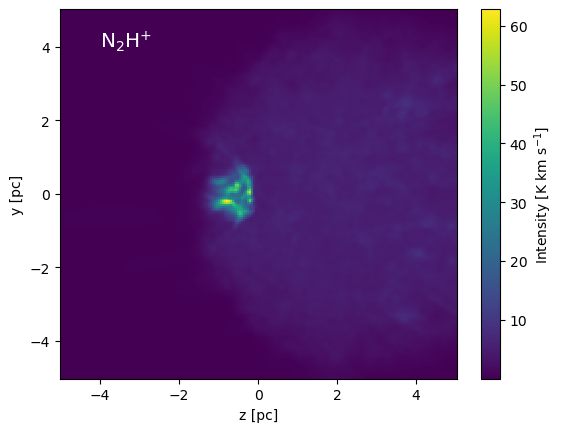}
    \includegraphics[width=0.3\textwidth]{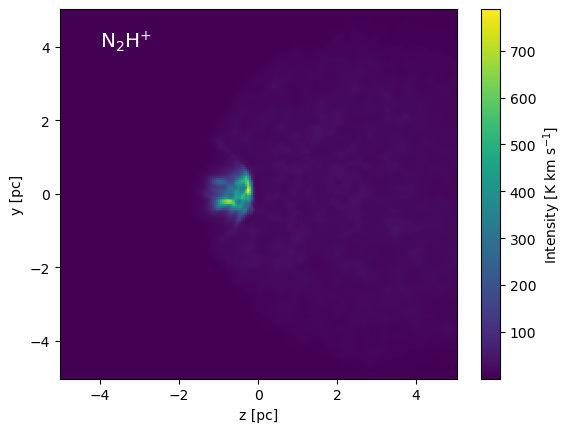}
\caption{Velocity Integrated Plots, Point of View is perpendicular to axis of collision \\ Left Column: Milky Way Disc - Center Column: CMZ - Right Column: CMZ with high $\zeta$}
\label{fig:vInt}
\end{figure*}
We select $10^5$ SPH particles which end the simulation within $5 \pc$ of the collision centre, and post-process their chemical evolution using {\sc uclchem} \citep{holdship2017}, a time-dependent gas-grain chemical code using the UMIST12 network \citep{mcelroy2013}. We take elemental abundances from \citet{sembach2000}, and assume an initially-atomic composition for all species except hydrogen, which is in the form of H$_2$. The molecular abundances for each tracer particle are then evolved from these initial conditions up to the final time, using the full evolutionary history of the particle's density and temperature. The background ultraviolet (UV) radiation field is taken to be $1.7$ \citet{habing1968} units, and the cosmic ray ionisation rate (CRIR) is set to $1.3 \times 10^{-17} {\rm \, s^{-1}}$. Although these parameters do not enter into the SPH simulations, the increased gas temperature of clouds in the CMZ is likely to be caused by an enhanced CRIR in this region \citep[e.g.][]{clark2013}, so we additionally investigate running the chemistry for the CMZ model with a CRIR increased by a factor of 100. To estimate shielding column densities to attenuate the UV field, we follow the approach in \citet{priestley2021} and multiply the local volume density by the Jeans length, and convert to extinction via a factor of $6 \times 10^{-22} {\rm \, mag \, cm^2}$ \citep{bohlin1978}.

 In addition to an increased CRIR, the UV background in the CMZ may also be higher than the Solar neighbourhood value adopted here \citep[e.g.][]{winter2020,henshaw2023}. However, the high column densities of our CMZ model clouds ($\sim 50$ magnitudes of extinction in the cloud centres) mean that this is unlikely to have any significant effect on the bridging features which are the focus of this paper, as they arise from the high-density compressed region at the collision interface. \citet{clark2013} investigated CMZ clouds with enhanced UV background and a self-consistent calculation of the shielding column densities using the TreeCol algorithm \citep{clark2012b}. In these simulations, the effect of the higher UV field is mostly limited to a very thin region around the cloud edges, suggesting that it is unlikely to have much impact on the molecular line emission we are interested in here.

Finally, we obtain position-position-velocity (PPV) cubes of the molecular line emission using the radiative transfer code {\sc radmc3d} \citep{dullemond2012}. We use {\sc splash} \citep{price2007} to interpolate physical properties from the SPH simulation onto a cube with side length $10 \pc$ and spatial resolution $0.05 \pc$, and assign each cell the molecular abundances of the nearest tracer particle. Our output spectra cover a velocity range of $\pm 12 \kms$ with a resolution of $0.12 \kms$. We investigate the $J=1-0$ transitions of CO, HCO$^+$ and N$_2$H$^+$, and the $J=2-1$ transition of CS, using collisional data taken from the LAMDA database \citep{schoier2005}. Dust properties are a mixture of \citet{ossenkopf1994} and \citet{mathis1983} opacities, as described in \citet{clark2012}.

\section{Results}
\label{sec:results}
\noindent The left-most column of Fig.~\ref{fig:yAvg} displays the Milky Way disc cloud collision model in position-velocity space. The normalised line profiles, averaged over the central $0.5$ pc, are shown in Fig.~\ref{fig:profile}. In this environment, \( \co \) is highly abundant at densities between $10^3-10^5 \pcc$ (Fig.~\ref{fig:envAbun}), and so emits strongly in both the molecular clouds themselves and the bridging feature of the collision (Fig.~\ref{fig:yAvg}).  \( \cs \), \( \hco \), \( \nth \) on the other hand are far less abundant and have higher critical excitation densities \citep{shirley2015}, so are most visible in emission in the compression region at the interface between the clouds (Fig.~\ref{fig:vInt}).  In position-velocity space (Fig.~\ref{fig:yAvg}), the clouds themselves are either only weakly apparent or completely absent, making the identification of bridging features challenging. The intensity contrast between the broad bridge feature and the unperturbed clouds (Fig.~\ref{fig:profile}) makes the signature weak in CS and almost invisible in HCO$^+$. N$_2$H$^+$ displays a reversed feature (only the denser cloud and the bridge region are visible), but at such low absolute intensities (Fig.~\ref{fig:yAvg}) that this would not be practically observable.

In the CMZ, the increased temperature and density alter the emission properties of \( \co \) such that the ambient cloud material becomes much brighter.  This serves to make \( \co \) bright everywhere and reduces the contrast of the bridging feature (Fig.~\ref{fig:yAvg}), thus limiting its viability as a tracer molecule in the CMZ environment.  On the other hand, \( \cs \) increases in abundance, and emits more readily due to the higher density and temperature throughout all densities (Fig.~\ref{fig:envAbun}).  In the CMZ \( \cs \) shows the bridging feature with a much clearer intensity contrast to the cloud than \( \co \) (Fig.~\ref{fig:profile}).  The abundance curves for \( \hco \) and \( \nth \) flatten in the CMZ, losing the peak around $10^{4}$ cm$^{-3}$ in the Milky Way disc, and as such they no longer preferentially trace the dense region of the collision.

In the CMZ with a high ionization rate, \( \co \) again has a low contrast between cloud and bridge regions.  \( \cs \), however, decreases in abundance back down to levels similar to the Milky Way Disc (Fig.~\ref{fig:envAbun}).  The image of the bridging feature in the \( \cs \) in Fig.~\ref{fig:yAvg} is less pronounced.  However, the high ionization rate causes the \( \hco \) abundance curve to mimic the abundance curve of \( \co \) in the Milky Way Disc (Fig.~\ref{fig:envAbun}).  As such, the position-velocity graph of \( \hco \) provides a readily apparent bridging feature in the highly ionized CMZ (Fig.~\ref{fig:yAvg}).  Similarly, \( \nth \) provides a strong bridging feature, albeit with a greater emphasis on the the feature itself.  In the case of a high ionization rate, \( \hco \) and \( \nth \) may be more promising tracers of cloud collisions than \( \co \) and \( \cs \); their line profiles (Fig.~\ref{fig:profile}) are morphologically similar to that of CO in the Milky Way Disc, whereas the bridge feature itself is significantly less apparent in CO.

\section{Discussion}
\label{sec:discussion}
\noindent Some recent studies into CMZ cloud collisions have used \( \cs \) as an alternative cloud collision tracer to \( \co \)  \citep[e.g.][]{busch2022}.  Based on the results of this study, that choice appears to be a valid one for identifying cloud collisions in the CMZ, which would not be the case in the Milky Way disc.  Studies in the CMZ using \( \co \) have encountered difficulties in identifying strong bridging features \citep[e.g.][]{gramze2023}, in agreement with our results. While we have focused on the $J=1-0$ transition, we find qualitatively indistinguishable results for the CO $J=3-2$ line, although sufficiently high-$J$ transitions with critical densities comparable to the other lines investigated here may recover the broad bridge feature in the CMZ. The extreme environment of the CMZ appears to allow the use of other species which under normal conditions preferentially trace dense gas, in an analogous manner to \( \co \) as signatures of cloud collisions.  Similarly, the presence of widespread supersonic turbulence in the CMZ \citep{henshaw2023} may make SiO, normally a shock tracer, more suited to tracing the bulk of the cloud material, as in \citet{armijos2020}.

The widespread \( \nth \) emission seen in the high-CRIR CMZ model is in good agreement with the observed behaviour of this molecule \citep{santamaria2021}, and in start contrast to its role as a tracer of high-density gas in the Milky Way disc \citep{pety2017,kauffmann2017,tafalla2021}.  To our knowledge, this is the first time this behaviour has been reproduced using a self-consistent treatment of the dynamics and chemistry; \citet{petkova2023} use a fixed \( \nth \) abundance estimated from CMZ observations, while \citet{hsu2023} include an increased CRIR in their chemical model, but do not investigate the higher densities and temperatures present in the CMZ.  In addition to its role as a tracer of dense gas, the altered chemical behaviour in the CMZ is likely to affect the relationship between the \( \nth \) emission and the star formation rate proposed by \citet{priestley2023b} under Milky Way disc conditions.  Simulations with more realistic thermodynamics than the isothermal equation of state adopted here \citep[e.g.][Cusack et al. in prep.]{clark2013} will be required to investigate this point in more detail.

\section{Conclusion}
\label{sec:conclusion}
\noindent We have examined the simulated behavior of molecular line emission in the Milky Way Disc, and the CMZ with a normal and enhanced cosmic ray ionization rate.  We investigate the utility of these lines for investigating the presence of cloud-cloud collisions in  different environments.  We have concluded the following:
\begin{enumerate}
    \item Low-$J$ \( \co \) lines, while a good tracer molecule for Milky Way Disc collisions, fail to provide adequate contrast in the CMZ to properly identify bridging features.
    \item \( \cs \) emits more strongly under CMZ conditions than in the Milky Way disc, as such it provides a much more pronounced bridging feature.
    \item For CMZ simulations with an enhanced ionization rate, the visibility of the bridging feature in both \( \co \) and \( \cs \) is reduced, but \( \hco \) and \( \nth \) both provide very strong collisions signatures due to their enhanced chemical abundances and the higher gas densities and temperatures found in the CMZ.
\end{enumerate}

\section*{Acknowledgements}
 We thank the anonymous referee for a useful report on an earlier version of this paper. FDP acknowledges the support of a consolidated grant (ST/W000830/1) from the UK Science and Technology Facilities Council (STFC).

\bibliographystyle{mnras}
\bibliography{bibliography}

\end{document}